\documentclass[reprint,aps,prx,showpacs,showkeys,groupedaddress,longbibliography]{revtex4-1}

\usepackage{graphicx}        
\usepackage{grffile}         
\usepackage{graphics}
\usepackage{epsfig}
\usepackage{verbatim}
\usepackage{dcolumn}        
\usepackage{bm}            
\usepackage{amsmath}
\usepackage{mathtools}
\usepackage{physics}
\usepackage{hyperref}        
\hypersetup{colorlinks,breaklinks,
            linkcolor=black,urlcolor=black,
            anchorcolor=black,citecolor=black}
\usepackage{color}
\usepackage[dvipsnames]{xcolor}
\usepackage{xspace}
\usepackage{float}
\usepackage{textcomp}
\newcommand{\EQ}[1]{Eq.~(\ref{eq:#1})}
\newcommand{\EQS}[2]{Eqs.~(\ref{eq:#1}) and (\ref{eq:#2})}
\newcommand{\FIG}[1]{Fig.~\ref{fig:#1}}

\def\bal#1\eal{\begin{align}#1\end{align}}

\begin{document}
\title{Anomalous thermal relaxation of Langevin particles in a piecewise-constant potential}

\author{Matthew Walker}
\affiliation{Department of Physics, University of Virginia, Charlottesville, VA 22904, USA}
\author{Marija Vucelja}
\affiliation{Department of Physics, University of Virginia, Charlottesville, VA 22904, USA}
\email{mvucelja@virginia.edu}

\begin{abstract}
We consider the thermal relaxation of a particle in a piecewise-constant potential landscape subject to thermal fluctuations in the overdamped limit. We study the connection between the occurrence of the Mpemba effect, the presence of metastable states, and phase transitions as a function of the potential. We find that the Mpemba effect exists even in cases without metastable states. In the considered physical system, the borders of the areas where the effect happens correspond to either eigenvector changes of direction or to phase transitions. Finally, we discuss the topological aspects of the strong Mpemba effect and propose using topology to search for the Mpemba effect in a physical system. 
\end{abstract}

\pacs{}
\keywords{Thermal relaxation, Nonequilibrium Statistical Physics, Mpemba effect, Thermal quench, Optimal cooling, Optimal heating, overdamped-Langevin equation}

\maketitle
\section{Introduction}
The interest in anomalous relaxation phenomena stems from deepening our basic knowledge and understanding of the dynamics of systems out of equilibrium. Equally important are pragmatic efforts to utilize anomalous relaxations to optimize heating and cooling processes in metallurgy, provide better sample preparation, material design, and develop efficient numerical samplers. 

The Mpemba effect is an anomalous relaxation phenomenon in which a system starting at a hot temperature cools down faster than an identical system starting at an initially lower temperature when both are coupled to an even colder bath. The effect was originally observed in water~\cite{Mpemba}, where the proposed explanations include effects of the presence of dissolved gasses and solids in water that affects its cooling properties~\cite{Katz}, convection and evaporation~\cite{VYNNYCKY2015243,Vynnycky2010,2006jeng}, supercooling~\cite{95Auerbach} and reorientation of hydrogen bonds~\cite{C4CP03669G}. Besides water the Mpemba effect was experimentally observed in colloidal systems~\cite{Kumar2020aa,kumar2021anomalous}, polymers~\cite{18Mpembapoly}, magnetic alloys~\cite{2010arXiv1011.3598C}, and clathrate-hydrates~\cite{Ahn2016gf}. It was simulated in granular fluids~\cite{PhysRevE.99.060901,2017Prados}, spin glasses~\cite{Baity-Jesi15350}, quantum systems~\cite{2019arXiv190512029N}, nanotube resonators~\cite{Greaney2011}, cold gasses~\cite{Keller_2018}, mean-field antiferromagnets~\cite{Mpemba17VRHK}, ferromagnets~\cite{D1CP00879J}, systems without equipartition~\cite{Gijon2019br}, molecular dynamics of water molecules~\cite{Jin2015aa}, molecular binary mixtures~\cite{doi:10.1063/5.0050530} and driven granular gasses~\cite{PhysRevE.102.012906}. Some of the recent theoretical advances include the formulation of the Mpemba effect for a general system~\cite{Lu16052017}, the definition of the strong Mpemba effect and its topological properties~\cite{Mpemba17VRHK}, the observation that optimal heating strategy may include pre-cooling the sample~\cite{PhysRevLett.124.060602}, and the notion that in the case of metastability, the Mpemba effect corresponds to a non-monotonic temperature dependence of extractable work~\cite{10.3389/fphy.2021.654271}. 

Motivated by expanding our intuition on the Mpemba effect, we search for the phenomenon in the case of a Langevin particle diffusing and advecting on a potential energy landscape in the overdamped limit. We study the Mpemba effect, or non-monotonic thermal relaxation, as a function of parameters defining the potential landscape. We should note that such a system was experimentally studied by Kumar and Bechhoefer in~\cite{Kumar2020aa}. They used optical tweezers to create a double-well potential and then watched how a colloidal particle submerged in water relaxes to equilibrium. They were the first to observe the strong Mpemba effect in experiment~\cite{Kumar2020aa}. The same group, together with Ch\'etrite, was also the first to see the inverse Mpemba effect~\cite{kumar2021anomalous}. Here we theoretically consider several simple potentials and focus on the salient features when one observes the strong Mpemba effect as a function of the potential. 

An overly simplistic heuristic explanation attempt of the Mpemba effect is that the "colder" system is stuck in metastable states, compared to the identical system starting from a "hotter" temperature. The heuristic suggests that metastable states of the right kind of geometry are necessary for the effect to happen. However, below we show that metastable states are not required for the Mpemba effect to occur. More concretely, in the case of a piecewise-constant potential, we analytically and numerically find phase space regions where the Mpemba effect exists, and those include areas without metastable states. In our model, we observe that the borders of the regions where the effect happens correspond to eigenvector changes of direction and phase transitions. We discuss how to look for anomalous relaxation behavior and how to exploit it in applications.  

The paper is organized as follows. In section~\ref{sec:model} we introduce the physical model, section~\ref{sec:mpemba} defines the Mpemba effect and relevant topological properties of it. In section~\ref{sec:potential} we specify the potential and solve for the Mpemba effect. Section~\ref{sec:gen-remarks} contains our main results on the strong Mpemba effect in the case of overdamped-Langevin dynamics in a piecewise-constant potential and~\ref{sec:summary} summarizes the paper. 

\section{Model}
\label{sec:model}
We consider a particle subject to potential $\tilde{U}(\tilde{x})$, and damping $\tilde{\gamma}$, in a thermal environment, characterized by noise $\tilde{\Gamma}(\tilde{t})$. The mean and the variance of the noise are
    \begin{align}
    \langle  \tilde{\Gamma}(\tilde{t})\rangle = 0 \text{ and }
    \langle \tilde{\Gamma}(\tilde{t})\tilde{\Gamma} (\tilde{t}') \rangle = 2 \tilde{\gamma} k_B \tilde{T}_b \delta (\tilde{t} - \tilde{t}'),
\end{align}
where $\tilde{T}_b$ is the temperature of the surrounding heat bath and $k_B$ is the Boltzmann's constant. For damping $\tilde{\gamma}$ large compared to inertia, the motion of the particle is described by the overdamped-Langevin equation
\begin{align}
    \frac{d\tilde{x}}{d\tilde{t}}+  \frac{1}{\tilde{\gamma}}\frac{d \tilde{U}}{d\tilde{x}} = \frac{\tilde{\Gamma}(\tilde{t})}{\tilde{\gamma}}. 
\end{align}
The evolution of a probability density $\tilde{p}(\tilde{x},\tilde{t})$ of having a particle at position $\tilde{x}$ at time $\tilde{t}$ obeys the Fokker-Planck equation
\begin{align}
\frac{\partial \tilde{p}}{\partial \tilde{t}} = \frac{\partial}{\partial \tilde{x}}\left[\frac{1}{\tilde{\gamma}}\frac{d\tilde{U}}{d\tilde{x}}\tilde{p}\right] + \frac{2 k_B\tilde{T}_b}{2\tilde{\gamma}}\frac{\partial^2\tilde{p}}{\partial \tilde{x}^2},
\end{align}
c.f.~\cite{ZinnJustin,Gardiner,Risken}. The Fokker-Planck equation arises in many situations, such as in Brownian motion~\cite{Risken,KRAMERS1940284}, colloids held with optical tweezers~\cite{Kumar2020aa}, chemical reactions~\cite{van_Kampen,Gardiner}, fluctuations of the current on a Josephson junction, and stretching of a polymer~\cite{Engel,SMS,Bird}.

It is convenient to use the following normalized coordinate $x$, time $t$, potential $U$ and temperature $T$ defined as 
\begin{align}
\label{eq:units}
    x \equiv \frac{2 \pi}{L}\tilde {x},\, 
    t \equiv \frac{(2 \pi)^2}{L^2}\frac{k_B \tilde{T}_b}{\tilde{\gamma}}\tilde{t},\,
    U \equiv \frac{\tilde{U}}{k_B \tilde{T}_b},\,
    T \equiv \frac{\tilde{T}}{\tilde{T_b}}. 
\end{align}
The normalized coordinate is in the domain $x\in \mathcal{D}\equiv[-\pi,\pi]$. Note that the normalized potential $U$ and time $t$ depend on the bath temperature $\tilde{T}_b$. In the new variables the Fokker-Planck equation is 
\begin{align}
    \frac{\partial p}{\partial t} &= \mathcal{L}_F\, p  = - \frac{\partial J}{\partial x},
\end{align}
where $\mathcal{L}_F$ is the Fokker-Planck operator
\begin{align}
    \mathcal{L}_F&\equiv \frac{\partial}{\partial x} U' + \frac{\partial^2}{\partial x^2},
\end{align}
and $J(x,t)$ is the probability current 
\begin{align}
      J(x,t) &\equiv -e^{-U(x)}[e^{U(x)} p(x,t)]'. 
\end{align}
Here $U' \equiv dU/dx$, and the equilibrium probability density at $T_b = 1$ is 
\begin{align}
    &\pi(x|T=1) = \frac{e^{- U(x)/T}}{Z(T)}\bigg\vert_{T=1}, 
\end{align}
where $Z(T) \equiv \int _{\mathcal{D}} \pi (x|T)\, dx$ is the norm. The Fokker-Planck operator $\mathcal{L}_F$ is not self-adjoint, but it can be transformed into a self-adjoint operator $\mathcal{L}$ with the following transformation
\begin{align}
    \mathcal{L} = e^{\frac{U(x)}{2}}\mathcal{L}_F e^{-\frac{U(x)}{2}} = \frac{\partial^2}{\partial x^2} - V(x),
\end{align}
where 
\begin{align}
    V(x) \equiv e^{\frac{U(x)}{2}}\left(\frac{\partial^2}{\partial x^2 }e^{-\frac{U(x)}{2}}\right) = \frac{U'\,^2}{4} - \frac{U''}{2}, 
\end{align}
for details see e.g.~\cite{Risken}. Finding the spectrum of the Fokker-Planck operator $\mathcal{L}_{F}$, reduces to solving a Schr\"odinger eigenvalue problem 
\begin{align}
\label{eq:Sch-eq}
    \mathcal{L} \psi_\mu = \lambda_\mu \psi _ \mu. 
\end{align}
The eigenvalues are ordered and non-positive $\lambda_1 = 0 > \lambda _2 \geq \lambda _3 \geq ...$ The general solution with the initial condition $p(x',0)$ is 
\begin{align}
p(x,t) = \int _\mathcal{D}  G(x,x',t) p(x',0)\,d x', 
\end{align}
where the transition probability is 
\begin{align}
G(x,x',t) = e^{-\frac{U(x)}{2}+\frac{U(x')}{2}}\sum _{\mu} \psi _\mu(x) \psi _\mu ^* (x') e^{-|\lambda_\mu| t},
\end{align}
and the eigenvectors $\psi_\mu$ fulfil the completeness relation $\sum _\mu \psi _\mu(x) \psi _\mu ^*(x') = \delta (x - x')$. The first eigenvector, corresponding to $\lambda_1 =0$, is $\psi_1 (x) = e^{-U(x)/2}/\sqrt{Z(1)}$. Thus the general solution for the probability density is 
\begin{align}
    \label{eq:prob-evo}
    p(x,t) &= \frac{e^{-U(x)}}{Z(1)} + \sum _{\mu > 1} a_\mu e^{-\frac{U(x)}{2}}\psi_\mu(x)e^{-|\lambda_\mu|t},
    \end{align}
    with
\begin{align}
    \label{eq:a_mu-coef}
    a_\mu &\equiv \int _\mathcal{D}dx' p(x',0) e^{\frac{U(x')}{2}} \psi_\mu ^* (x')\,dx'.
\end{align}
Assuming $\lambda_2 > \lambda_3$ at times $t \gg |\lambda_3|^{-1}$ we have
\begin{align}
    p(x,t) &\approx \frac{e^{-U(x)}}{Z(1)} +  a_2 e^{-\frac{U(x)}{2}}\psi_2(x)e^{-|\lambda_2|t}.
\end{align}

\section{The Mpemba effect}
\label{sec:mpemba}
Let us choose for the initial condition the equilibrium distribution at temperature $T$, i.e. 
\begin{align}
    p(x,0) = \pi(x|T) =\frac{e^{-U(x)/T}}{Z(T)}.
\end{align}
In this case the overlap coefficients $a_\mu$ are 
\begin{align}
\label{eq:acoefs}
    a_\mu(T) = Z^{-1}(T)\int _\mathcal{D}dx'  e^{-U(x')\left(\frac{1}{T}- \frac{1}{2}\right)} \psi_\mu (x')\,dx'. 
\end{align}
Notice that because of orthogonality of $\psi_\mu$ eigenvectors we get $a_\mu (1) = 0$ as expected (no cooling or heating if $T = T_b = 1$). As $T \to \infty$, $a_\mu$ becomes independent of temperature, and plateaus to a constant. Assuming that $\lambda_2 > \lambda_3$ the Mpemba effect occurs for $a_2(T)$ non-monotonic as a function of initial temperature $T$~\cite{Lu16052017}. The strong Mpemba effect occurs for $a_2(T) = 0$ for select $T\neq T_b$,~\cite{Mpemba17VRHK}. If $a_2(T) = 0$ for all $T$, then the relaxation to equilibrium does not have that mode and one needs to look at $\mu > 2$ for anomalous relaxations. 

\subsection{Parity analysis}
One way to check for the strong Mpemba effect in cooling is to check for parity, 
\begin{align}
\label{eq:Pdir} 
&\mathcal{P}_{\rm dir} \equiv \left[-\left.\frac{d a_2}{d T}\right|_{T=1} a_2 (T = \infty)\right],
\\
\label{eq:Pinv}
&\mathcal{P}_{\rm inv} \equiv \lim _{\varepsilon\to 0^+}\left[\left.\frac{d a_2}{d T}\right|_{T=1} a_2 (\varepsilon)\right],
\end{align}
which was introduced in~\cite{Mpemba17VRHK}. There is an odd number of zero crossing of $a_2(T)$ between $T\in (1,\infty)$ if $\mathcal{P}_{\rm dir} > 0$. From \EQ{a_mu-coef} we have 
\begin{align}
    a_2 (\infty) &=\frac{1}{2 \pi}\int _\mathcal{D} \psi _2 (x) e^{\frac{U(x)}{2}}\, dx,
\\
\left.\frac{d a_2}{d T}\right|_{T=1}
&=\frac{Z(2)}{Z(1)}\left[ \langle U\psi_2\rangle_1 - \langle U\rangle _1 \langle \psi_2\rangle_2 \right],
\end{align}
where $\langle g(x) \rangle _T \equiv \int _\mathcal{D} g(x)e^{-\frac{U(x)}{T}} [Z(T)]^{-1}\,dx$. Thus the parities of the direct and inverse strong Mpemba effect are
\begin{align}
\mathcal{P}_{\rm dir}
 =&\left[(
 \langle U\rangle _1 \langle \psi_2\rangle_2- \langle U\psi_2\rangle_1) \int _\mathcal{D} e^{\frac{U(x)}{2}} \psi_2(x) dx\right],
 \\
\nonumber
\mathcal{P}_{\rm inv} =& \lim _{\varepsilon\to 0^+}\bigg[\left(
 \langle U\psi_2\rangle_1-\langle U\rangle _1 \langle \psi_2\rangle_2\right)
  \times
  \\
  &\quad\quad\quad\times 
\int _\mathcal{D} e^{\frac{U(x)}{2}-\frac{U(x)}{\varepsilon}} \psi_2(x)dx\bigg].
\end{align}

\section{piecewise-constant potential}
\label{sec:potential}
We investigate the existence of the Mpemba effect for simple potentials to gain intuition when the effect occurs. In the case of symmetric potentials $V(x)$ and $U(x)$ eigenvector of the first excited state, $\psi_2$, is odd. Over symmetric domains, the overlap coefficient $a_2$, given by \EQ{acoefs}, is automatically zero, and thus there is no Mpemba effect related to this overlap coefficient. For more details see~\ref{sec:symmetric-potential}. Similarly, for the case of the quadratic potential $U(x) = k x^2/2$, which corresponds to the Ornstein-\"Uhlenbeck process, we show in~\ref{sec:quadratic-potential}, that there is no Mpemba effect. 

As the next case in simplicity -- below, we introduce an analytically solvable case of a piecewise-constant potential with three different regions and derive analytically and numerically the regions in the phase space defined by the potential parameters where the system displays the strong Mpemba effect. Section~\ref{sec:gen-remarks} contains our main results. 

Let us choose the potential as  
\begin{align}
\label{eq:piecewise-potential}
    U(x) = \begin{cases} 
    U_1, &x \in [-\pi, -\alpha\pi/2)\\ 
    U_0, &x\in [-\alpha\pi/2, \pi/2]\\
    0, &x\in(\pi/2,\pi]
    \end{cases},
\end{align}
where $U_0$, $U_1$, and $\alpha\in[0,1]$. Our potential has finite jumps at $-\alpha \pi/2$ and $\pi/2$, and it diverges to infinity at $\pm \pi$. For a finite discontinuity of the potential the probability current must be constant to satisfy the conservation of probability. Assuming we have a finite jump at $x$ the "jump" conditions are 
\begin{align}
    & \label{eq:jump1}
    e^{\frac{U(x^+)}{2}}\psi _\mu (x^+) = e^{\frac{U(x^-)}{2}}\psi _\mu (x^-) ,
    \\ 
    \nonumber
    &e^{-\frac{U(x^+)}{2}}\left[\psi '_\mu (x^+) + \frac{1}{2} U'(x^+)\psi_\mu(x^+)\right] =
    \\  \label{eq:jump2}
    &e^{-\frac{U(x^-)}{2}}\left[\psi '_\mu (x^-) + \frac{1}{2} U'(x^-)\psi_\mu(x^-)\right].
\end{align}
We assume that the potential has a positive infinite value at the edges of the domain. At the edges of the domain, the probability current must be zero, $J(\pm \pi)=0$, if we are to have a non-trivial steady state.
\begin{figure}
\includegraphics[width=0.85\columnwidth]{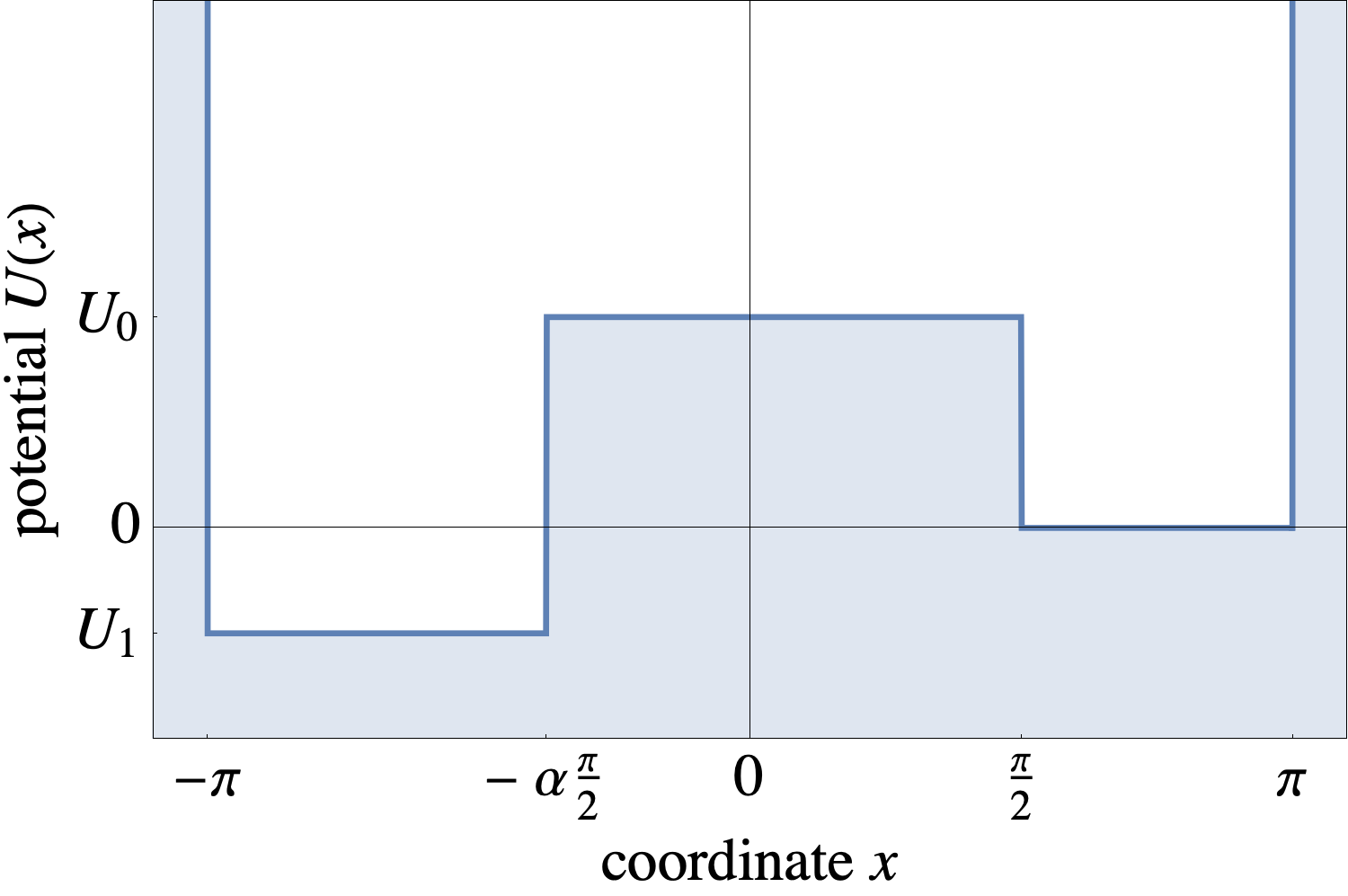}
\caption{\label{fig:fig-piecewise-const-alpha-diff-U2-0-potential-v01.png}Piecewise-constant potential $U(x)$ with parameters $U_0$, $U_1$, and $\alpha\in [0,1]$.}
\end{figure}

\subsection{Bistable symmetric rectangular potential well}
Let us choose $U_1 = 0$, $\alpha = 1$, and vary $U_0$. The potential in this case corresponds to a bistable symmetric rectangular well. The Fokker-Planck equation is analytically solvable,~\cite{1979ZPhyBRisken}. The eigenvector of the first excited state is 
\begin{align}
    \psi_2(x) &= \begin{cases}
     -\frac{1}{\sqrt{\pi}}\cos[\nu (\pi + x)],& x \in [-\pi, -\frac{\pi}{2})\\
    \frac{1}{\sqrt{\pi}}\sin[\nu x],&|x| \leq \frac{\pi}{2}\\
    \frac{1}{\sqrt{\pi}}\cos[\nu (\pi - x)],& x \in (\frac{\pi}{2},\pi]
    \end{cases},
\end{align}
with $\nu\equiv\frac{2}{\pi}\arctan[e^{-U_0/2}]$. The relevant eigenvalues are nondegenerate: $\lambda_1 = 0$, $\lambda_2 = \nu^2$, and $\lambda_3 = 1$. The first excited state $\psi_2$ is odd; thus $a_2 = 0$, as an integral of an odd function in a symmetric domain. Therefore there is no Mpemba effect associated with $a_2$. The result is consistent with the finding of Kumar and Bechhoefer, who experimentally saw that there is no Mpemba effect associated with $a_2$ for their double-well symmetric potential in a symmetric domain~\cite{Kumar2020aa}.

\subsection{Varying the heights and the widths of a piecewise-constant potential}
Let us now consider the cases of $\alpha \in [0,1]$, and vary $U_0$ and $U_1$. The eigenfunctions are 
\begin{align}
    \label{eq:psi}
    \psi _\mu=\begin{cases}
   A_\mu\cos [\sqrt{\lambda_\mu}(x + \pi)], &  -\pi \leq x <  \frac{-\alpha\pi}{2}
    \\
    B_\mu \cos [\sqrt{\lambda_\mu}x] + C_\mu \sin[\sqrt{\lambda_\mu}x], &  \frac{-\alpha\pi}{2}\leq x\leq \frac{\pi}{2}
    \\
    D_\mu\cos [\sqrt{\lambda_\mu}(x - \pi)], &\frac{\pi}{2}< x\leq \pi
    \end{cases}.
\end{align}
The zero-current boundary conditions, $\psi'_\mu(\pm \pi) = 0$, are fulfilled by construction. The jump conditions, \EQS{jump1}{jump2}, and the normalization of $\psi_\mu$'s, specify the coefficients $A_\mu$, $B_\mu$, $C_\mu$ and $D_\mu$. The transcendental equation that specifies $\lambda_2$ is 
\begin{align}
\nonumber
    &\frac{-e^{U_1}\cos\left[\frac{\sqrt{\lambda_2}\alpha \pi}{2}\right]+e^{U_0}\sin\left[ \frac{ \sqrt{\lambda_2} \alpha \pi}{2} \right]\tan\left[ \sqrt{\lambda_2}\pi \left( 1 - \frac{\alpha}{2} \right) \right]}{e^{U_1}\cos\left[{\frac{\sqrt{\lambda_2}\alpha \pi}{2}}\right]-e^{U_0}\sin\left[ \frac{\sqrt{\lambda_2} \alpha \pi}{2}  \right]\tan\left[ \sqrt{\lambda_2} \pi \left(1 - \frac{\alpha}{2} \right) \right]}
\\ \label{eq:transcendental-eq}
   &=\frac{\cot \left[ \frac{\sqrt{\lambda_2} \pi}{2} \right] -e^{U_0}\tan \left[ \frac{\sqrt{\lambda_2 \pi}}{2} \right]}{e^{U_0}+1}.
\end{align}
Note that the width parameter $\alpha$ appears only inside trigonometric functions, and thus its contribution is bounded.
For general $\alpha$, $\lambda_2$ cannot be found in an explicit form. Such form however exists in the case of $\alpha = 1$, and $\alpha = 0$. Below we present analytic results in the two cases and numerical results for arbitrary width parameter $\alpha$. 

\subsubsection{Equal widths of the left and right sections, the $\alpha = 1$ case}
\begin{figure}
    \centering
    \includegraphics[width=0.9\columnwidth]{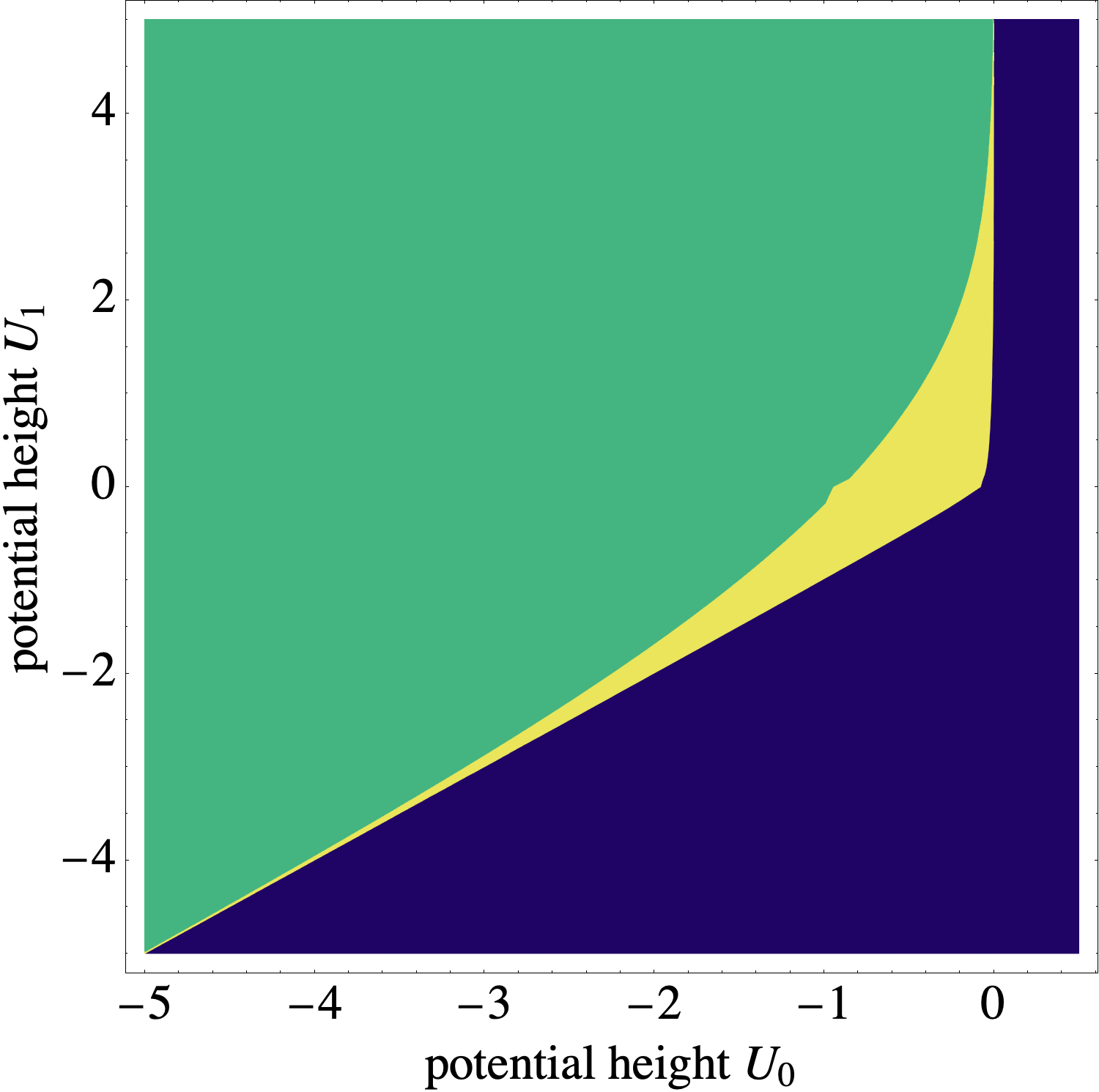}
    \caption{\label{fig:fig-piecewise-const-alpha-1-U2-0-parity-inv-dir-region-plot-v01.png}The strong Mpemba effect for $\alpha = 1$. In the green region we have the direct strong Mpemba effect (the parity $\mathcal{P}_{\rm dir} > 1$) and in the yellow region we have computed the inverse strong Mpemba effect (the parity $\mathcal{P}_{\rm inv}>1$, where $\varepsilon = 0.02$). The parities are defined in~\EQS{Pdir}{Pinv}. In the blue region there is no strong Mpemba effect. Here $T_b = 1$ and $k_B = 1$.}
\end{figure}

In the case $\alpha = 1$ we have the transcendental equation gives $\lambda_2$ as 
\begin{align}
\lambda_2 = \left[\frac{2}{\pi}\tan^{-1}\left[\sqrt{\frac{2-\tanh
   \left[\frac{U_0}{2}\right]-\tanh \left[\frac{U_0-U_1}{2}\right]}{2+\tanh
   \left[\frac{U_0}{2}\right]+\tanh
   \left[\frac{U_0-U_1}{2}\right]}}\right]\right]^2.
\end{align}
Plugging in $\psi_2$ and $\lambda_2$, into \EQ{acoefs} we get the overlap coefficient $a_2$
\begin{align}
   \nonumber
   a_2=& \frac{2 \sin \left[\frac{\pi  \sqrt{\lambda_2}}{2}\right]}{\pi 
   \sqrt{\lambda_2}}\times 
   \\
   &\times\frac{
   \left(A_2e^{\frac{U_0}{T}+\frac{U_1}{2}}+2 B_2
   e^{\frac{U_0}{2}+\frac{U_1}{T}}+D_2
   e^{\frac{U_0}{T}+\frac{U_1}{T}}\right)}{
   \left(e^{\frac{U_0}{T}+\frac{U_1}{T}}+e^{\frac{U_0}{T}}+2 e^{\frac{U_1}{T}}\right)}.
\end{align}
The jump conditions, \EQS{jump1}{jump2}, and the normalization of $\psi_2$, specify the coefficients $A_2$, $B_2$, $C_2$ and $D_2$. The zeros of the numerator of $a_2$ define the set of temperatures for which we have the strong Mpemba effect~\cite{Mpemba17VRHK}.
\begin{figure}
    \centering
    \includegraphics[width=\columnwidth]{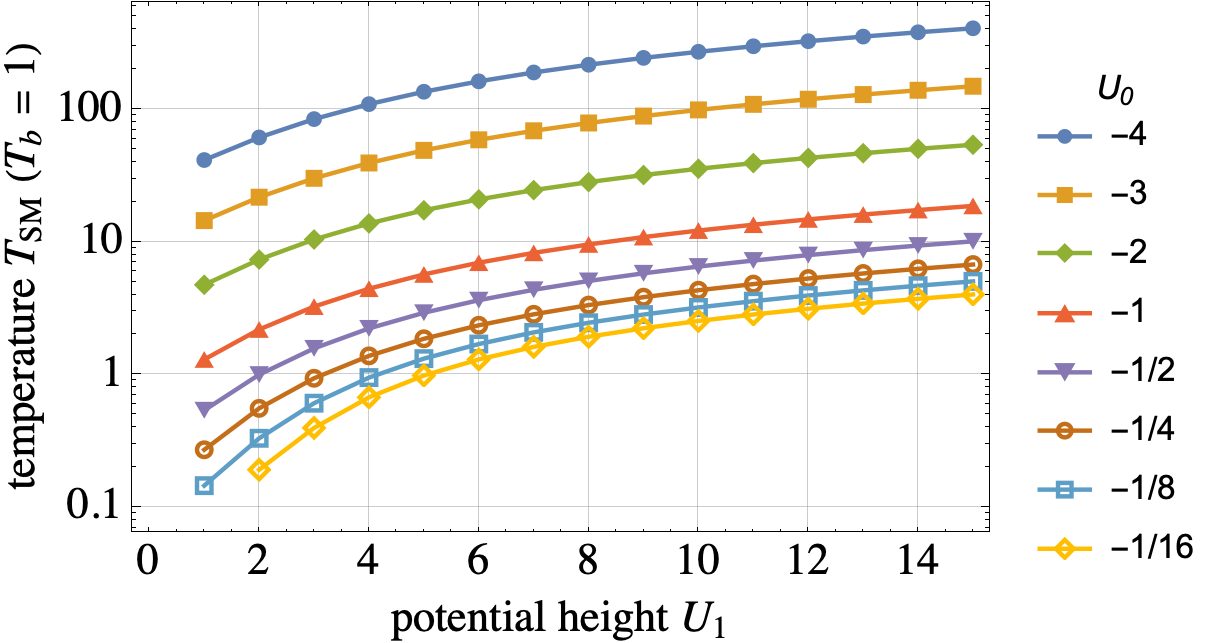}
    \caption{\label{fig:fig-piecewise-const-alpha-1-U2-0-zeros-a2-T_SM-v01.png}The temperature of the strong Mpemba effect $T_{\rm SM}$ as a function of potential parameters $U_0$, $U_1$ and $\alpha = 1$. Here $k_B = 1$ and $T_b = 1$.}
\end{figure}
\begin{figure}
\includegraphics[width=\columnwidth]{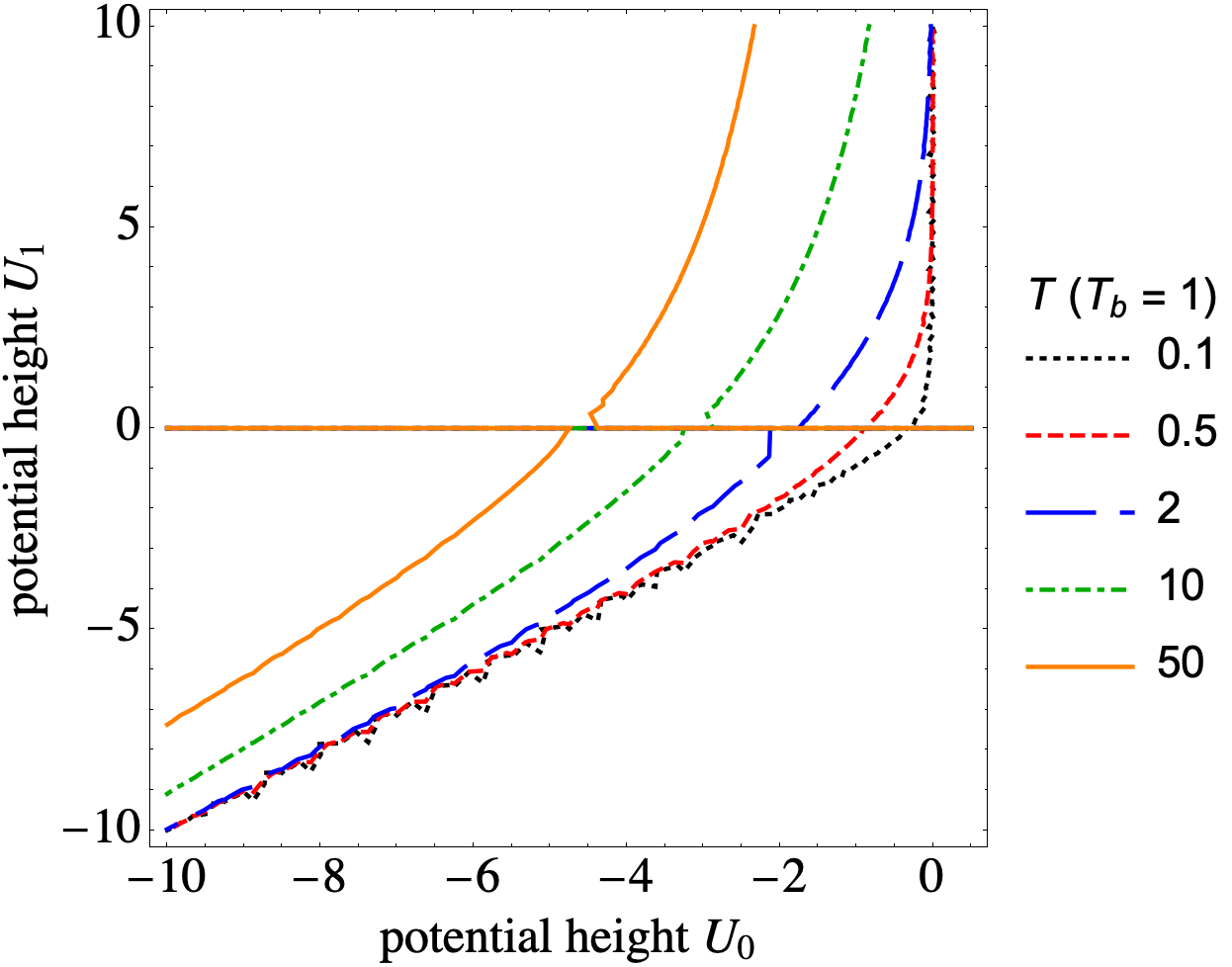}
\caption{\label{fig:fig-piecewise-const-alpha-1-U2-0-zeros-a2-T-diff-v01.png}The strong Mpemba effect is present along the isolines of $a_2 = 0$ for the potential heights $U_1$, $U_0$, $\alpha = 1$ and initial temperature $T$. Here $T_b = 1$ and $k_B =1$.}
\end{figure}
For particular choices of potential parameters $U_0$ and $U_1$, we get the Mpemba effect. The strong Mpemba temperature $T_{\rm SM}$ as a function of $U_0$, and $U_1$ is shown on~\FIG{fig-piecewise-const-alpha-1-U2-0-zeros-a2-T_SM-v01.png}. The isolines of the strong Mpemba effect in the $U_0U_1-$plane are depicted on~\FIG{fig-piecewise-const-alpha-1-U2-0-zeros-a2-T-diff-v01.png}. On \FIG{fig-piecewise-const-alpha-1-U2-0-parity-inv-dir-region-plot-v01.png} the green region shows the region of existence of the direct strong Mpemba effect (cooling), and the yellow region shows the region of existence of inverse strong Mpemba (heating). In the blue region, there is no strong Mpemba effect. We observe the strong Mpemba effect for $U_1 > U_0$ and $U_0 < 0$, which corresponds to the absence of metastable states. Note that we see the Mpemba effect in the absence of metastable states -- this challenges the heuristic explanation attempt described in the introduction, c.f. also~\cite{Kumar2021qa}. 

Below in Section~\ref{sec:gen-remarks} we argue that the strong Mpemba effect for $\alpha = 1$ happens when the mismatch between the initial probability and the final probability in the left region matches that the mismatch between the initial and final probabilities of the right region.   

\subsubsection{Wide left section, the $\alpha=0$ case}
\begin{figure}
    \centering
    \includegraphics[width=0.9\columnwidth]{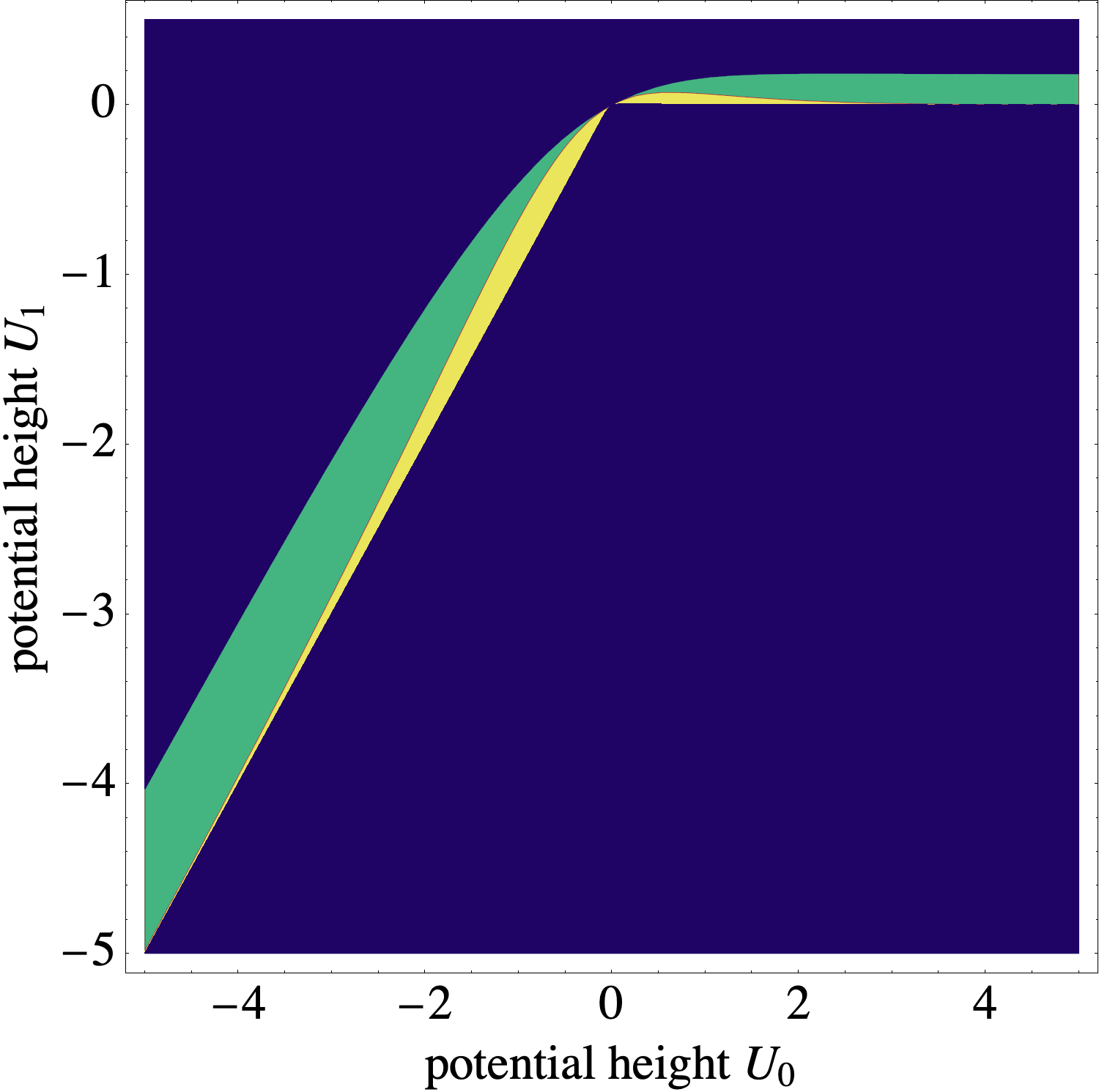}
    \caption{\label{fig:fig-piecewise-const-alpha-0-U2-0-parity-inv-dir-region-plot-v01.png}The strong Mpemba effect as a function of the potential parameters $U_0$, $U_1$, and $\alpha = 0$. In the green region we have the direct strong Mpemba effect (the parity for the direct effect is $\mathcal{P}_{\rm dir} > 1$; see~\EQ{Pdir}) and in the yellow region we have the inverse strong Mpemba effect (the parity for the inverse effect is $\mathcal{P}_{\rm inv}>1$) region. The parity $\mathcal{P}_{\rm inv}$ was computed by choosing $\varepsilon = 0.02$ in~\EQ{Pinv}. In the blue region there is no strong Mpemba effect. Here $T_b = 1$ and $k_B = 1$.}
\end{figure}
\begin{figure}
    \centering
    \includegraphics[width=\columnwidth]{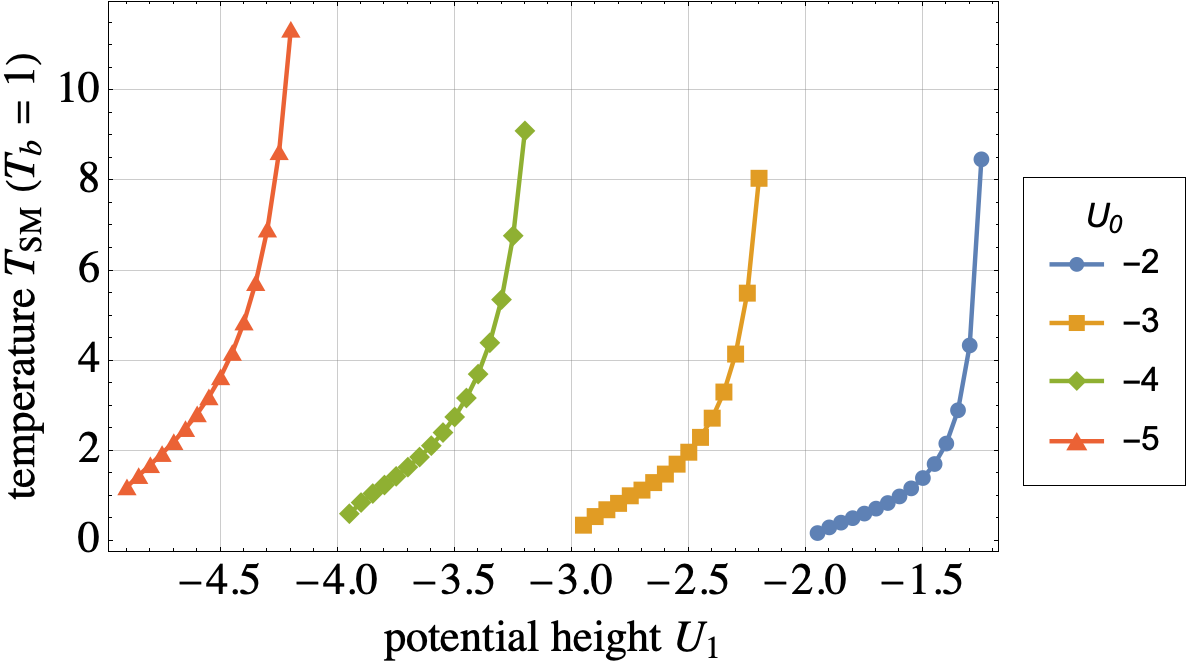}
    \caption{\label{fig:fig-piecewise-const-alpha-0-U2-0-zeros-a2-T_SM-v01.png}The temperature of the strong Mpemba effect $T_{\rm SM}$ as a function of potential parameters $U_0$, $U_1$, and $\alpha = 0$. Here $T_b = 1$ and $k_B = 1$.}
\end{figure}
\begin{figure}
\includegraphics[width=0.8\columnwidth]{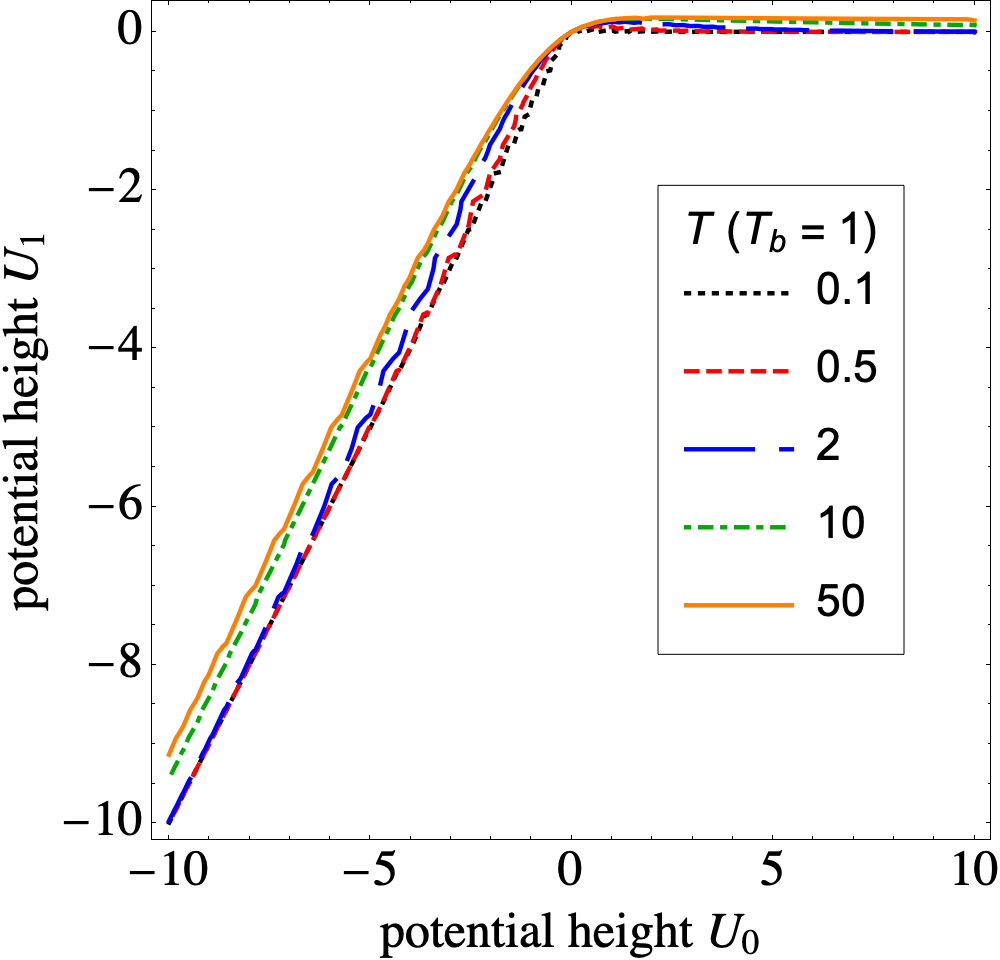}
\caption{\label{fig:fig-piecewise-const-alpha-0-U2-0-zeros-a2-T-diff-v01.png}The strong Mpemba effect is present along the isolines of $a_2 = 0$ for the potential parameters $U_1$, $U_0$, $\alpha = 0$ and initial temperature $T$. Here $T_b = 1$ and $k_B = 1$.}
\end{figure}
Above we demonstrated the $\alpha=1$ case is exactly solvable. Next we obtain an analytic solution for the $\alpha=0$ case. In this example, the width of the left section is twice the width of the center section and right section. The form of the eigenfunctions is~\EQ{psi}, but the eigenvalue $\lambda_2$ is different 
\begin{align}
\lambda_2 = \left[\frac{2}{\pi} \tan ^{-1}\left(\sqrt{\frac{e^{U_1-U_0}-\tanh
   \left[\frac{U_0}{2}\right]+1}{e^{U_1-U_0}+\tanh
   \left[\frac{U_0}{2}\right]+1}}\right)\right]^2,
\end{align}
and the domains with the strong effect are changed respectively, see~\FIG{fig-piecewise-const-alpha-0-U2-0-parity-inv-dir-region-plot-v01.png}. Now we see that the region with the strong Mpemba effect is dramatically smaller. It requires fine-tuning the potential to demonstrate the Mpemba effect. However, unlike the $\alpha=1$ case, one now has a Mpemba effect for a barrier in the middle section ($U_0 > U_1$ and $U_0>0$) and metastable states, akin in the experiment of Kumar and Bechhoefer~\cite{Kumar2020aa}. The strong Mpemba temperature $T_{\rm SM}$ as a function of $U_0$, and $U_1$ is shown on~\FIG{fig-piecewise-const-alpha-0-U2-0-zeros-a2-T_SM-v01.png}. The isolines of the strong Mpemba effect in on the $U_0U_1-$plane are depicted on~\FIG{fig-piecewise-const-alpha-0-U2-0-zeros-a2-T-diff-v01.png}.

\subsubsection{Varying middle section's width, the case $\alpha \in(0,1)$}
\begin{figure}
    \centering
    \includegraphics[width=\columnwidth]{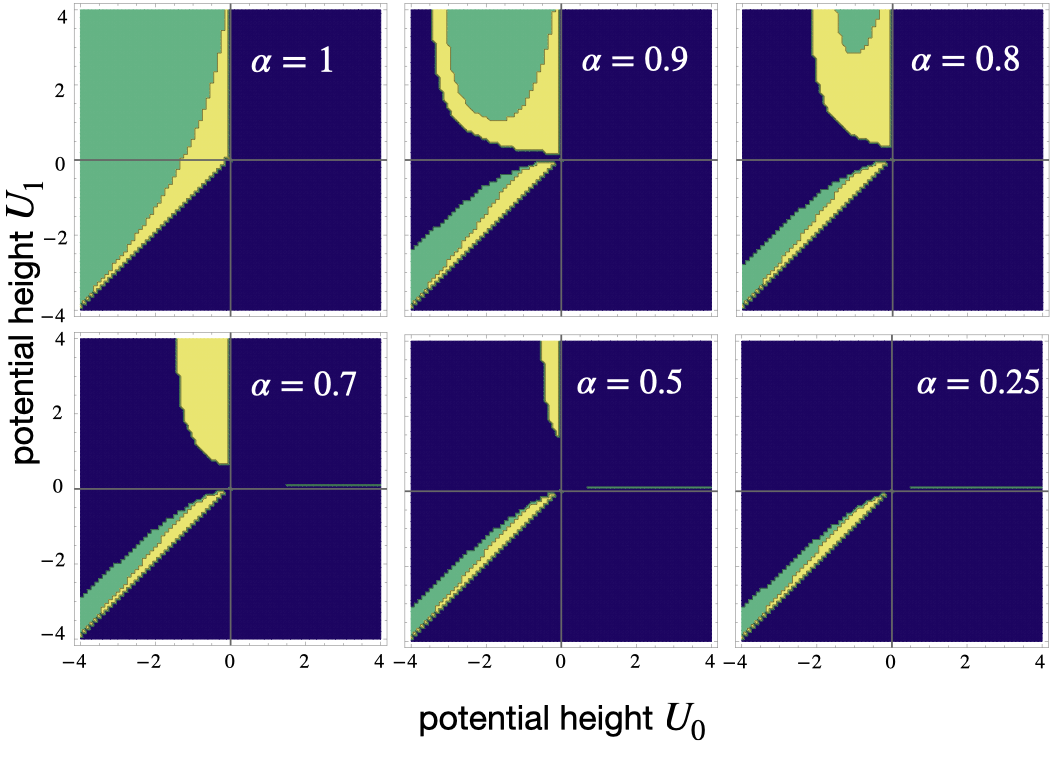}
    \caption{\label{fig:fig-piecewise-const-alpha-diff-U2-0-parity-inv-dir-num-plot-v02.001.png}The strong Mpemba effect as a function of potential parameters $U_0$, $U_1$, and $\alpha \in[0.25,1]$. In the green region we have the direct strong Mpemba effect (the parity $\mathcal{P}_{\rm dir} > 1$) and in the yellow region we have the inverse strong Mpemba effect ($\mathcal{P}_{\rm inv}>1$) region. There is no strong Mpemba effect in the blue region. Here we looked for the strong effect between initial temperatures $0.02\, T_b \leq T\leq 200\, T_b$, where $T_b = 1$ and $k_B = 1$. The parity was calculated via~\EQS{Pdir}{Pinv}.}
\end{figure} 
Next, we consider what happens if we change the width of the left and middle piecewise sections, with $\alpha \in (0,1)$. It is important to note that we are not solving the system perturbatively; we are solving the whole problem for new widths, starting with the transcendental equation given by~\EQ{transcendental-eq}. 
 In the case of arbitrary $\alpha$ \EQ{transcendental-eq} does not have an explicit solution for $\lambda_2$, but it is solvable numerically. After the eigenvalue is obtained, the coefficients $A_2$ $B_2$, $C_2$ and $D_2$ are calculated from the jump conditions, \EQS{jump1}{jump2}, and the normalization of the eigenvector. Now we can go about calculating $a_2$ numerically and study what happens. In the parity plots~\FIG{fig-piecewise-const-alpha-diff-U2-0-parity-inv-dir-num-plot-v02.001.png} we see the behavior changes immediately. This change can be understood through the symmetry breaking of the middle section. The eigenvector for this region is, $ B_\mu \cos [\sqrt{\lambda_\mu}x] + C_\mu \sin[\sqrt{\lambda_\mu}x]$. When we integrate this eigenvector over a symmetric domain, as we do in the $\alpha=1$ case, the contribution of the $\sin[\sqrt{\lambda_\mu x}]$ piece always vanishes. By changing $\alpha$, we break this symmetry and now $\sin[\sqrt{\lambda_\mu}x]$ term in the middle section, $-\alpha \pi / 2 \leq x \leq \pi/2$, will also contribute to the overlap $a_2$.

\subsection{General remarks on the strong Mpemba effect for the piecewise-constant potential} 
\label{sec:gen-remarks}

\subsubsection{Regions of the direct and strong Mpemba effect}

Here the direct and inverse strong Mpemba effect regions are disjoint, see~\FIG{fig-piecewise-const-alpha-diff-U2-0-parity-inv-dir-num-plot-v02.001.png}, while in general, the effects can coexist. For example, in Glauber dynamics on the mean-field antiferromagnet on a complete bipartite graph, there is a region where one has both strong Mpemba effects~\cite{Mpemba17VRHK}.

Also, note that the region where we have the inverse effect in this range of parameters seems smaller than where we have the direct effect. It results from a temperature unit scale we have imposed on the problem by setting $T_b = 1$. Namely, there is less "room" to create non-zero curvature between the $T_b$ and zero temperature, then between $T_b$ and infinity, which corresponds to less phase space area for the inverse strong Mpemba effect than the direct strong Mpemba effect.

\subsubsection{Ratio of the mismatch in equilibrium probabilities in the flanking regions}

To shed some intuition on when we see the strong Mpemba effect we look at the difference of the equilibrium probabilities for the particle to be at the left and the right region at the bath temperature $T_b$ and the temperature of the strong Mpemba effect $T_{\rm SM}$. The equilibrium probability of a particle being in region $\mathcal{D}_i$ is
\begin{align}
    \label{eq:eq-prob-region}
    \Pi_{i}(T) \equiv \int _{\mathcal{D}_i} \pi (x|T) dx,
\end{align}
where $\mathcal{D}_1 = [-\pi, -\alpha\pi/2)$ is the left, $\mathcal{D}_0 = [-\alpha\pi/2,\pi/2]$ is the middle and 
$\mathcal{D}_2 = (\pi/2,\pi]$ is the right region.
The ratio of the difference in equilibrium probabilities is defined as 
\begin{align}
    R \equiv \frac{\Pi _1 (T_b) - \Pi_1(T_{\rm SM})}{\Pi _2 (T_b) - \Pi_2(T_{\rm SM})}.
\end{align}
\begin{figure}
    \centering
    \includegraphics[width=\columnwidth
    ]{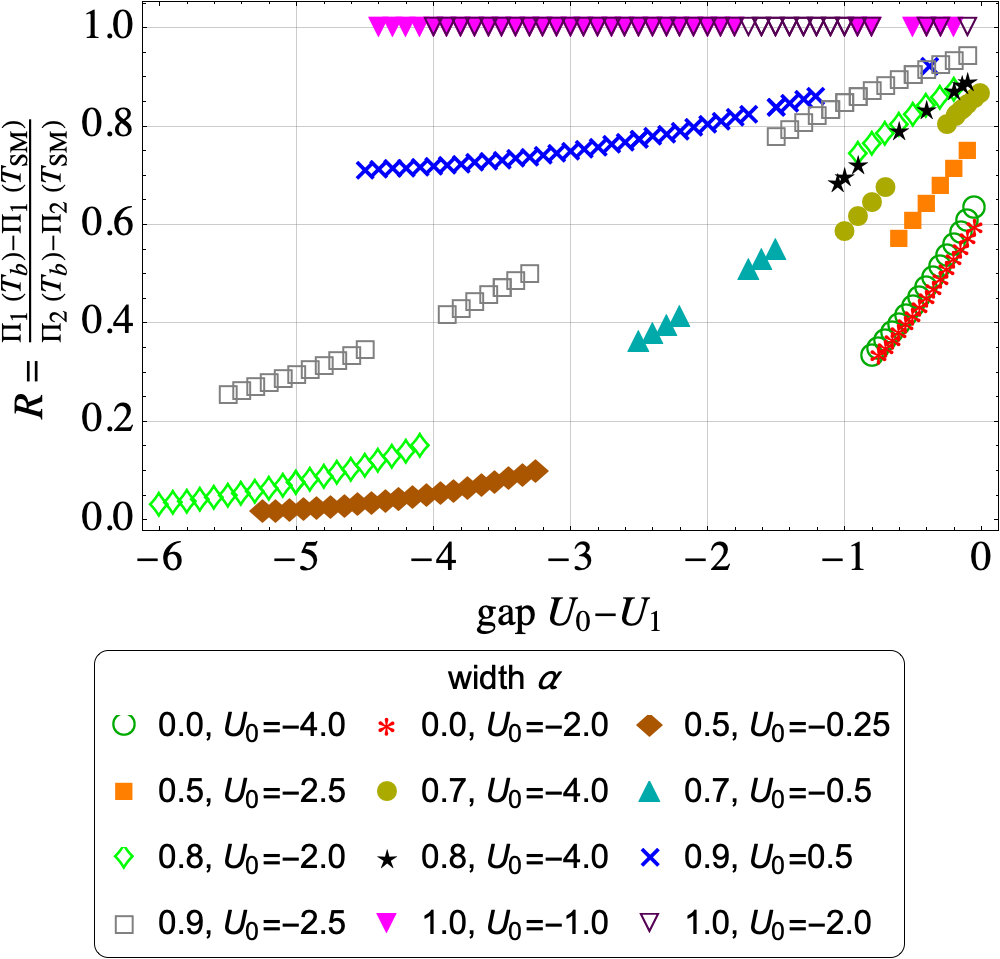}
    \caption{Ratio of difference of equilibrium probabilities at the bath temperature, $T_b$, and the temperature where we have the strong Mpemba effect, $T_{\rm SM}$, for the left region (1) and the right region (2) as a function of the gap $U_0 - U_1$, $U_0$ and $\alpha$. Here $T_b = 1$ and $k_B = 1$. We notice that the ratio $R$ is equal to 1 for $\alpha = 1$. In other cases, $\alpha \in [0,1)$, the ratio depends on both the gap $U_0 - U_1$ and $U_0$.}
    \label{fig:fig-piecewise-const-alpha-diff-U2-0-ratio-cum-prob-gap-v01.png}
\end{figure}

From~\FIG{fig-piecewise-const-alpha-diff-U2-0-ratio-cum-prob-gap-v01.png} we notice that for left and right regions of the same width, $\alpha = 1$ case, the ratio $R = 1$. In this case, we have the strong Mpemba effect only if there is a difference between the initial and final probabilities in the left region, matching that of the right region. Also, $R = 1$, can be used as an implicit formula for $T_{\rm SM}$. 

For flanking regions of different widths, $\alpha < 1$, the ratio $R$ is less than $R(\alpha = 1) = 1$. I.e., in this case, we have the strong Mpemba effect when the wider region contains less probability mismatch than the narrower region -- how much less depends on all of the parameters of the potential, that is $R(U_0 - U_1, U_0, \alpha)$. Namely, we see from~\FIG{fig-piecewise-const-alpha-diff-U2-0-ratio-cum-prob-gap-v01.png} that the ratio $R$ is a function of both the gap $U_0 - U_1$ and $U_0$. As we make the left region wider, reduce $\alpha$, the dependence on $U_0$ becomes weaker compared to the dependence on the gap $U_0 - U_1$.

Note that in the case of the metastable Mpemba effect, described in~\cite{kumar2021anomalous, Kumar2020aa}, the authors see the effect for potentials that simultaneously satisfy $\Pi _1 (T_b) = \Pi_1(T_{\rm SM})$ and $\Pi _2 (T_b) =\Pi_2(T_{\rm SM})$, which is quite different from our case. Indeed, for the piecewise-constant potential that we are considering, metastability is not needed to have the effect. Even more, for $\alpha = 1$, we do not have the effect if we have metastability.

\subsubsection{Topological considerations}
The existence of the strong Mpemba effect could be thought of as a topological invariant~\cite{Mpemba17VRHK}. Namely, it is a non-trivial intersection of the locus of points corresponding to the equilibrium distribution at different temperatures and the $a_2=0$ hyperplane. The number of times this locus of points intersects the $a_2=0$ hyperplane is the intersection number and was named the Mpemba index by the authors of~\cite{Mpemba17VRHK}. As a topological invariant, the Mpemba index can change under perturbations, but its modulo two cannot. Our results show agreement with this assertion. In our analysis of the piecewise-constant potential, we show that the strong Mpemba effect cannot be removed or introduced without changing the Mpemba index modulo two, which can only happen if, as laid out in~\cite{Mpemba17VRHK}: 
\begin{itemize}
    \item[(i)] The perturbation changes the ordering of the eigenvalues -- it causes $\lambda_3$ to become larger than $\lambda_2$.
    \item[(ii)] The perturbation causes $a_2(0)$ or $a_2(\infty)$ or both to change sign. For this to occur, the eigenvector $\psi_2$ must change "direction." 
    \item[(iii)] There is a phase transition. For example, the ground state of the system changes.
\end{itemize}
We obtain the full spectrum of eigenvalues analytically and conclude that eigenvalues $\lambda_2$ and $\lambda_3$ do not cross in our case; thus, (i) never happens. In our case, removing or introducing the strong Mpemba effect requires that the system goes through a change of the direction of the eigenvector (ii) or through a phase transition (iii), or both. 

\FIG{fig-piecewise-const-alpha-1-U2-0-parity-inv-dir-region-plot-v01.png},~\FIG{fig-piecewise-const-alpha-0-U2-0-parity-inv-dir-region-plot-v01.png}, and~\FIG{fig-piecewise-const-alpha-diff-U2-0-parity-inv-dir-num-plot-v02.001.png} provide a simple phase diagrams. The green, yellow, and blue regions are divided by domain walls, demarking the region of existence of the direct, the inverse strong Mpemba effect, and the absence of both effects, respectively. 

For equally wide outer sections, in the $\alpha=1$ case, one can only get a strong Mpemba on a part $U_0<0$ half-plane where $U_0 <U_1$ (see \FIG{fig-piecewise-const-alpha-1-U2-0-parity-inv-dir-region-plot-v01.png}). In this case, one cannot get a strong Mpemba effect in $a_2$ if the middle section is a barrier. Regardless of how small one makes the middle section, i.e.~$U_0$, it cannot be the highest potential height. The symmetry of the problem protects this. It seems that as if one needs remove the metastable states for the effect to occur. Likewise, choosing $U_0<0$ and crossing the $U_1 = U_0$ line toward $U_1 < U_0$ introduces a metastable state and removes the strong Mpemba effect. However, note that simply "removing" metastable states will not introduce the effect; in the region $U_1<U_0<0$, there is no strong Mpemba effect, despite the absence of metastable states. 

In the case that the outer potential sections have different widths, the $\alpha \neq 1$ case, there now exists additional domain walls, compared to the $\alpha = 1$ case, where the Mpemba index modulo two can change, see~\FIG{fig-piecewise-const-alpha-0-U2-0-parity-inv-dir-region-plot-v01.png} and~\FIG{fig-piecewise-const-alpha-diff-U2-0-parity-inv-dir-num-plot-v02.001.png}. As before, these domain walls correspond to the eigenvector changing the direction and to changes of the ground state. 

The line between the direct and the inverse effect (between green and yellow regions on the phase diagrams on ~\FIG{fig-piecewise-const-alpha-1-U2-0-parity-inv-dir-region-plot-v01.png},~\FIG{fig-piecewise-const-alpha-0-U2-0-parity-inv-dir-region-plot-v01.png}, and~\FIG{fig-piecewise-const-alpha-diff-U2-0-parity-inv-dir-num-plot-v02.001.png}) corresponds to two zeros of $a_2$, one at $T > T_b$ and the other at $T = T_b$, merging into one at $T =T_b$ and the becoming two distinct zeros again where one is now at $T < T_b$ and the other remains at $T = T_b$.

To conclude, by studying how a Brownian particle diffuses on a potential energy landscape, we see how the particle behaves vastly differently depending on the geometry of the potential landscape. Intuitively this is to be expected, but what is interesting is that there are particular initial temperatures for which the system relaxes exponentially faster than when starting from other temperatures. By studying this phenomenon in our piecewise-constant potential, we see that this behavior is protected by symmetries present in our problem and is robust to perturbations. Together, these provide intuition on the dynamical behavior of our Brownian particle. The described exotic behavior could be considered a topological phase because the system's behavior is topologically protected against perturbations. 

Additionally, out of the three cases which change the Mpemba index, stated in~\ref{sec:gen-remarks}, the phase transitions and the crossing of eigenvalues are properties of the potential and bath only; they do not depend on the initial conditions, while as the eigenvector some changes of direction are significant for specific initial conditions. Thus one could use eigenvalue crossings and phase transitions to gauge the domains which might yield the Mpemba effect. Such explorations might be useful for experimental and numerical applications. 

\section{Summary}
\label{sec:summary}
We studied the occurrence of the Mpemba effect in several simple potentials. We show that there is no Mpemba effect for symmetric potentials in symmetric domains related to the first excited state. We further show that to find a Mpemba effect, one needs to go beyond a quadratic potential to polynomials of higher degrees or make the diffusion coefficient spatially dependent. 

Next, we solved analytically and numerically the case of a piecewise-constant potential with variable height and variable width sections. We analyzed the existence of the strong Mpemba effect as a function of the parameters of the potential and remarked on the topological aspects of the strong effect. In particular, we found that in the case of equal-width outer sections and a variable height of the sections, there seems to be no strong Mpemba effect if the system has metastable states. I.e., the middle section cannot be a barrier between the two wells. If the outer sections are not of equal width, this condition is relaxed, and we can also have the Mpemba effect with metastable states present.
In summary, we challenge the intuition that for the strong Mpemba effect, one needs metastable states. Instead, we demonstrate by our example that it sometimes becomes more challenging to have a Mpemba effect if the potential contains metastable states. The phase diagrams that we obtained show manifestly different relaxation behavior on every line that denotes a change of the deepest well. 

Moreover, in the case of equal-width outer sections we found that the strong Mpemba effect occurs when the ratio of the mismatch between initial and final probabilities in the two outer sections is equal. 

A particle diffusing in a potential landscape is a frequent effective description in phenomenological theories. For an arbitrary potential, the problem is not analytically tractable. We chose this conceptually simple situation to gain intuition on anomalous relaxation processes and nonmonotonicity in relaxation times. We looked at a piecewise-constant potential, where we could solve for the dynamics of the probability distribution function exactly. We analyzed the connection between the occurrence of the strong Mpemba effect and the parameters of the potential. Based on topological considerations, we have identified the domains in the phase space, formed by the potential parameters, where one might expect to see the effect. These are areas on whose boundaries where there is a phase change (in our case, the deepest well changes) or where the eigenvector changes the direction significantly compared to the initial condition. In our example, for the phase space parameters that we checked, the areas with the strong Mpemba effect seemed simply connected. Studying the topology of such regions would be an exciting future avenue of study.

Understanding better when the Mpemba effect occurs will enable us to design auxiliary potential traps, such as with electromagnetic fields or optical lattices, that could facilitate optimal cooling and heating of our system and allow better preparation of a system in a particular state. 

\section{Acknowledgements}
MV and MW acknowledge discussions with Oren Raz, Zhiyue Lu, Rapha\"el Chetrite, John Bechhoefer, Gregory Falkovich, Baruch Meerson, and Aaron Winn. This material is based upon work supported by the National Science Foundation under Grant No.~DMR-1944539.

\section{Appendix}
\subsection{Symmetric potentials}
\label{sec:symmetric-potential}
For symmetric potentials, $V(x) = V(-x)$, the reflection operator, the operator that flips $\psi_\mu(x) \to \psi_\mu(-x)$, commutes with the Schr\"odinger operator $\mathcal{L}$. Thus each non-degenerate eigenvector of $\mathcal{L}$ must also be an eigenvector of the reflection operator, which implies that each eigenvector must be either even or odd under the reflection, see e.g.~\cite{GriffithsQM}. The ground state having no nodes must be even, and the first excited state having one node must be odd. In the case that $U$ is symmetric and $V$ is symmetric and the domain is symmetric, we have that $a_2(T)=0$ for all $T$, as an integral of an odd function over a symmetric domain. Hence there is no Mpemba effect associated with $a_2$ in this case. The effect can be present at a higher order, i.e. for $a_\mu$ with $\mu > 2$~\cite{Kumar2021qa}.

\subsection{Quadratic potential}
\label{sec:quadratic-potential}
For the quadratic potential $U(x) = k x^2/2$, the eigenvectors and eigenvalues are known. The case corresponds to the Ornstein-\"Uhlenbeck process, see e.g.~\cite{Gardiner}, which is described by the following Fokker-Planck equation
\bal
\partial _t p(x,t) = \partial _x \left[k x p(x,t)\right] + \frac{D_b}{2}\partial_x^2 p(x,t)
\eal
where $D_b = 2k_B T_b$ is the diffusion coefficient. The left eigenfunctions $\varphi_n$ and corresponding eigenvalues are 
\bal
\varphi_n = \left(2^n n!\right)^{-1/2}H_n \left[x \sqrt{\frac{k}{2k_B T_b}}\right],\quad \lambda_n = nk, 
\eal
where $H_n$ are the Hermite polynomials. The stationary solution of the Fokker-Planck equation is
\bal
\pi(x|T_b) = \sqrt{\frac{k}{2\pi k_BT_b}}\exp\left[- \frac{k x^2}{2k_BT_b}\right], 
\eal
and the general solution for the probability distribution is
\bal
p(x,t) =& \sum _{n = 0} \sqrt{\frac{k}{2\pi k_B T_b}}e^{- \frac{k x^2}{2k_B T_b}}\varphi_n(x)e^{-n kt}A_n,
\eal
with overlap coefficients $A_n \equiv \int ^\infty _{-\infty} \varphi_n p(x,0)\,d x $. In the case of $p(x,0) = \pi(x|T)$ the coefficients $A_n$ can be found explicitly as 
\begin{align}
A_{2n}(T) = \sqrt{\frac{(2n)!}{2^{2n}}}\frac{1}{n!}\left(\frac{T}{T_b}-1\right)^{n},\, A_{2n+1}(T) = 0.  
\end{align}
Note the overlap coefficients are $k$ independent.
For finite temperatures, the coefficient $A_{2n}$ is zero only for $T = T_b$. Therefore there is no strong Mpemba effect for the Ornstein-\"Uhlenbeck process. Moreover, 
$\left(T/T_b-1\right)^{n}$
is a monotonic function of $T$, thus there is no weak Mpemba effect either for the Ornstein-\"Uhlenbeck process. 

The absence of the Mpemba effect is expected. Namely, starting from a Gaussian (Boltzmann distribution at temperature $T$) and evolving with a Gaussian kernel to get another Gaussian (Boltzmann distribution at temperature $T_b$), we are allowed to vary only the width of the Gaussian, there is no other variable to vary~\cite{MarijaOren}. Thus with polynomial potentials and spatially uniform diffusion coefficients, to find a Mpemba effect, we need to go beyond a quadratic potential to polynomial of higher degree or other functions.  

\bibliography{VuceljaBib.bib}
\end{document}